\documentclass{PoS}

\title{The observational signatures of high-redshift dark stars }

\ShortTitle{The observational signatures of high-redshift dark stars }

\author{\speaker{Erik Zackrisson}\\
        Department of Astronomy, Stockholm University, 10691 Stockholm, Sweden\\
        E-mail: \email{ez@astro.su.se}}

\abstract{The annihilation of dark matter particles in the centers of minihalos may lead to the formation of so-called dark stars, which are cooler, larger, more massive and potentially more long-lived than conventional population III stars. Here, we investigate the prospects of detecting high-redshift dark stars with both existing and upcoming telescopes. We find that individual dark stars with masses below $\sim 10^3\ M_\odot$ are intrinsically too faint even for the upcoming James Webb Space Telescope (JWST). However, by exploiting foreground galaxy clusters as gravitational telescopes, certain varieties of such dark stars should be within reach of the JWST at $z\approx 10$. If more massive dark stars are able to form, they may be detectable by JWST even in the absence of lensing. In fact, some of the supermassive ($\sim10^7\ M_\odot$) dark stars recently proposed are sufficiently bright at $z\approx 10$ to be detectable even with existing facilities, like the Hubble Space Telescope and 8-10 m telescopes on the ground. Finally, we argue that since the hottest dark stars ($T_\mathrm{eff}\gtrsim 30000$ K) can produce their own HII regions, they may be substantially brighter than what estimates based on stellar atmosphere spectra would suggest.}
\FullConference{Identification of Dark Matter 2010\\
		 July 26 - 30 2010\\
		 University of Montpellier 2, Montpellier, France}

\begin{document}

\section{Introduction}
It has recently been recognized that annihilation of dark matter in the form of Weakly Interacting Massive Particles (WIMPs; e.g.~the lightest supersymmetric or Kaluza-Klein particles, or an extra inert Higgs boson) may have generated a first population of stars with properties very different from the canonical population III \cite{Spolyar et al. a}.  Because the first stars are likely to form in the high-density central regions of minihalos, annihilation of dark matter into standard model particles could serve as an additional energy source alongside or instead of nuclear fusion within these objects. This leads to the formation of so-called dark stars, which are predicted to be cooler, larger, more massive and potentially longer-lived than conventional population III stars \cite{Spolyar et al. a,Iocco b,Freese et al. a,Iocco et al.,Yoon et al.,Taoso et al.,Natarajan et al.,Freese et al. b,Spolyar et al. b,Umeda et al.,Freese et al. c,Ripamonti et al.,Sivertsson & Gondolo}. 

A significant population of high-redshift dark stars could have important consequences for the formation of  intermediate and supermassive black holes \cite{Spolyar et al. b,Sandick et al.}, for the cosmic evolution of the pair-instability supernova rate \cite{Iocco c}, for the reionization history of the Universe \cite{Schleicher et al.} and the X-ray and infrared extragalactic backgrounds \cite{Schleicher et al.,Maurer et al.}. Effects such as these can be used to indirectly constrain the properties of dark stars, but no compelling evidence for or against a dark star population at high redshifts has so far emerged. 

\section{Direct detection of dark stars with the James Webb Space Telescope}
In a recent paper \cite{Zackrisson et al. b}, we explore the prospects of detecting dark stars with the upcoming James Webb Space Telescope (JWST, scheduled for launch in 2014). Using the TLUSTY \cite{Hubeny & Lanz} and MARCS \cite{Gustafsson et al.} stellar atmosphere models, we derive the JWST broadband fluxes of the Spolyar et al. \cite{Spolyar et al. b} $M\lesssim 10^3\ M_\odot$ dark star models\footnote{Both the rest-frame stellar atmosphere spectra and the JWST broadband fluxes (at various redshifts) of these dark stars are publicly available through the VizieR service at the CDS: http://cds.u-strasbg.fr/}. We find, that even though all of these dark stars are intrinsically too faint to be detectable at $z\approx 10$, gravitational lensing by a foreground galaxy cluster (with magnificication $\mu\sim 100$) can in principle lift some of these dark stars above the JWST detection threshold. Even though this takes care of the brightness issue, lensing may at the same time render the surface number densities of high-redshift dark stars too low for detection. To predict the expected number of detections behind a single galaxy cluster, we use the Trenti \& Stiavelli (\cite{Trenti & Stiavelli}) predictions for the cosmic star formation rate of population III stars in minihalos and adopt MACS J0717.5+3745 at $z=0.546$ as the primary target cluster (arguably the best lensing cluster currently available for studies of high-redshift objects \cite{Zitrin et al. a}). We find that a handful of $z\approx 10$ dark stars may in principle be detected in very deep JWST exposures ($\approx 30$ h per filter) of this cluster, but that this requires that the typical dark star lifetimes are long ($\gtrsim 10^7$ yr) and that the fraction of population III stars that go through a dark star phase is very high ($\gtrsim 0.1$). Even though these are admittedly very challenging observations, this may be the only way to directly detect $M\lesssim 10^3\ M_\odot$ dark stars at high redshifts in the foreseeable future. 

\section{Direct detection of supermassive dark stars with the Hubble Space Telescope}
Recently, Freese et al. \cite{Freese et al. c} speculated that dark stars might grow to become even more massive than in the Spolyar et al. (\cite{Spolyar et al. b}) models, eventually reaching masses of up to $10^7 M_\odot$. As is to be expected, such supermassive dark stars would be sufficiently bright to be seen at $z\approx 10$--15 with JWST, even without the lensing boost of a foreground galaxy cluster. While there are several unresolved issues concerning the fueling and stability of such extreme objects, the most massive ones are also strongly constrained by existing observations. Using TLUSTY stellar atmosphere models, we predict that $\sim 10^7 \ M_\odot$ dark stars should be sufficiently bright at $z\approx 10$ to be readily detected with the Hubble Space Telescope (HST) or 8--10 m class telescopes on the ground \cite{Zackrisson et al. c}. Based on the non-detection of $z\approx 10$ candidates at the relevant $H$-band fluxes ($H_\mathrm{AB}\approx 26$ mag) in current survey data, we find that $\sim 10^7 \ M_\odot$ dark stars must be very rare and/or short-lived to meet these observational constraints.

\section{Nebular emission from hot dark stars}
\begin{figure}
\includegraphics[width=.6\textwidth]{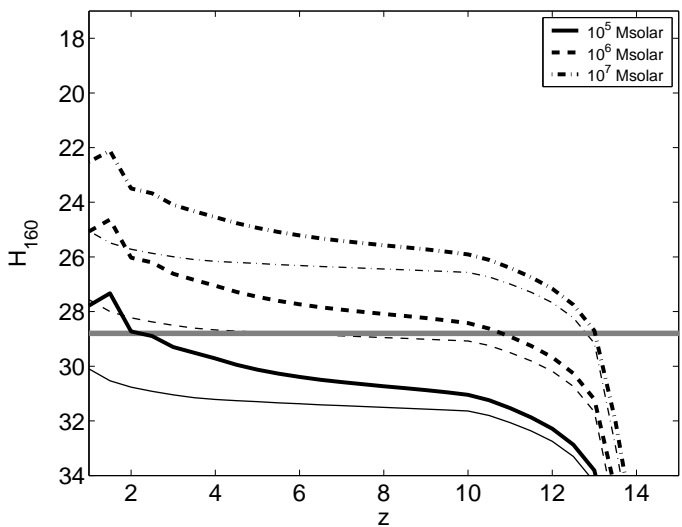}
\caption{The apparent $H_{160}$ AB magnitudes of the $10^5$--$10^7\ M_\odot$, $T_\mathrm{eff}= 51000$ K dark stars from Freese et al. (\cite{Freese et al. c}; their Table 4) as a function of redshift. Thin lines represent the models without nebular emission (i.e. fluxes based solely on TLUSTY stellar atmosphere spectra) whereas thick lines represent models with both stellar and nebular contributions. The horizontal gray line indicates the detection threshold of the deepest HST/WFC3 Hubble Ultra Deep Field observations currently available. The sharp drop in brightness at $z>13$ is due to absorption shortward of Ly$\alpha$ in the intergalactic medium. Due to the highly uncertain escape fraction of Ly$\alpha$ photons from dark stars, we have (somewhat conservatively) set the Ly$\alpha$ emission line flux to zero when estimating these $H_{160}$ fluxes. If the Ly$\alpha$ line were to be included, this would cause a substantial bump in the thick curves at $z=10.5$--14.}
\label{fig1}
\end{figure}

All current estimates of the fluxes of dark stars are based either on black body spectra \cite{Freese et al. c} or on stellar atmosphere models \cite{Zackrisson et al. b,Zackrisson et al. c}. However, the hottest ($T_\mathrm{eff}\gtrsim 30000$ K) dark stars may photoionize the gas in their host halos, thereby producing bright HII regions which could substantially boost the observed fluxes of these stars. The dynamical evolution of this gas is very complicated to predict in detail (see e.g. \cite{Kitayama et al.,Johnson et al.} for simulations relevant for conventional population III stars), and even if there initially is enough gas left in the halo to form an ionization-bounded nebula (and this is questionable in the case of supermassive dark stars), the gas may eventually be ejected from the halo. At that point, a huge, low surface brightness nebula will form in the intergalactic medium and is unlikely to contribute substantially to the fluxes of dark stars captured through aperture photometry in HST or JWST images. However, in the phase where the HII region remains compact and confined within the dark star halo, the observed broadband fluxes may be substantially boosted, in analogy with the case for young galaxies at high redshifts \cite{Zackrisson et al. d}.

In Fig.~\ref{fig1}, we estimate the flux boost due to nebular emission around the $10^5$--$10^7 M_\odot$, $T_\mathrm{eff}= 51000$ K dark stars predicted by Freese et al. (\cite{Freese et al. c}; their Table 4) by sending the TLUSTY stellar atmosphere spectra of these stars through the photoionization code Cloudy \cite{Ferland et al.}. These calculations assume a spherical, constant-density ($n(\mathrm{H})=10^2$ cm$^{-3}$), ionization-bounded nebula with no holes in the gas. This results in zero Lyman continuum leakage into the intergalactic medium, and therefore represents an {\it upper limit} on the likely flux contribution from nebular gas to the total fluxes of these dark stars -- both gas ejection from the halo and leakage through low-density regions (``holes'') in the nebula would render it lower. Nonetheless, this exercise shows that nebular emission may boost the $H$-band fluxes of these dark stars by up to $\approx 1$ mag at $z\approx 10$ and $\approx 2$ mag at lower redshifts. The expected colours of such dark stars will therefore also change due to the inclusion of nebular emission, just as in the case for young or star-forming galaxies (e.g. \cite{Zackrisson et al. a,Zackrisson et al. d}). Hence, nebular emission may play a crucial role in the observational pursuit for hot dark stars at high redshifts.

\section*{Acknowledgments}
Pat Scott, Claes-Erik Rydberg, Fabio Iocco, Bengt Edvardsson, G\"oran \"Ostlin, Sofia Sivertsson, Adi Zitrin, Tom Broadhurst, Paolo Gondolo, Garrelt Mellema, Ilian Iliev and Paul Shapiro are acknowledged for fruitful collaboration on dark stars and their observational signatures.


\begin{thebibliography}{99}
\bibitem{Spolyar et al. a}
Spolyar, D., Freese, K., \& Gondolo, P. 2008, {\it PhRvL}, 100, 051101
\bibitem{Iocco b}
Iocco, F. 2008, {\it ApJ}, 677, L1
\bibitem{Freese et al. a}
Freese, K, Bodenheimer, P., Spolyar, D., Gondolo, P. 2008, {\it ApJ}, 685, L101
\bibitem{Iocco et al.}
Iocco, F., Bressan, A., Ripamonti, E., Schneider, R., Ferrara, A., \& Marigo, P. 2008, {\it MNRAS}, 390, 1655
\bibitem{Yoon et al.}
Yoon, S.-C., Iocco, F., Akiyama, S. 2008, {\it ApJ}, 688, L1
\bibitem{Taoso et al.}
Taoso, M., Bertone, G., Meynet, G., \& Ekstr\"om, S. 2008, {\it PhRvD}, 78, l3510
\bibitem{Natarajan et al.}
Natarajan, A., Tan, J. C., \& O'Shea, B. W. 2009, {\it ApJ}, 692, 574
\bibitem{Freese et al. b}
Freese, K., Gondolo, P., Sellwood, J. A., \& Spolyar, D. 2009, {\it ApJ}, 693, 1563
\bibitem{Spolyar et al. b}
Spolyar, D., Bodenheimer, P., Freese, K., Gondolo, P. 2009, ApJ, 705, 1031 
\bibitem{Umeda et al.}
{Umeda}, H., {Yoshida}, N., {Nomoto}, K., {Tsuruta}, S., {Sasaki}, M., \& {Ohkubo}, T. 2009, {\it JCAP}, 08, 024
\bibitem{Freese et al. c}
Freese, K. Ilie, C., Spolyar, D., Valluri, M., Bodenheimer, P. 2010, {\it ApJ}, 716, 1397
\bibitem{Ripamonti et al.}
Ripamonti, E., Iocco, F., Bressan, A., Schneider, R., Ferrara, A, \& Marigo, P. 2010, {\it MNRAS}, 406, 2605 
\bibitem{Sivertsson & Gondolo}
Sivertsson, S., Gondolo, P. 2010, {\it ApJ}, submitted (arXiv1006.0025)
\bibitem{Sandick et al.}
Sandick, P., Diemand, J., Freese, K., Spolyar, D. 2010, arXiv1008.3552
\bibitem{Iocco c}
Iocco, F. 2009,   {\it Nucl. Phys. Proc. Suppl.}  194, 82 
\bibitem{Schleicher et al.}
Schleicher, D. R. G., Banerjee, R., \& Klessen, R. S. 2009, {\it PhRvD}, 79, 3510
\bibitem{Maurer et al.}
Maurer, A., et al. 2010, these proceedings
\bibitem{Zackrisson et al. b}
Zackrisson, E., Scott, P., Rydberg, C.-E., Iocco, F. Edvardsson, B., \"Ostlin, G., Sivertsson, S., Zitrin, A., Broadhurst, T., Gondolo, P. 2010, {\it ApJ} 717, 257 
\bibitem{Hubeny & Lanz}
Hubeny, I., \& Lanz, T. 1995, {\it ApJ}, 439, 875
\bibitem{Gustafsson et al.}
Gustafsson, B., Edvardsson, B., Eriksson, K., J\o{}rgensen, U. G., Nordlund, {\it A\&A}, \& Plez, B. 2008, A\&A 486, 951
\bibitem{Trenti & Stiavelli}
Trenti, M., \& Stiavelli, M. 2009, {\it ApJ}, 694, 879
\bibitem{Zitrin et al. a}
Zitrin, A., Broadhurst, T., Rephaeli, Y., Sadeh, S. 2009, {\it ApJ} 707, L102
\bibitem{Zackrisson et al. c}
Zackrisson, E., Scott, P., Rydberg, C.-E., Iocco, F., Sivertsson, S., \"Ostlin, G., Mellema, G., Iliev, I. T., Shapiro, P. R. 2010 {\it MNRAS}, 407, L74 
\bibitem{Kitayama et al.}
Kitayama, T., Yoshida, N., Susa, H., Umemura, M. 2004, {\it ApJ} 613, 631
\bibitem{Johnson et al.} 
Johnson, J. L., Greif, T. H., Bromm, V., Klessen, R. S., Ippolito, J. 2009, {\it MNRAS}, 399, 37
\bibitem{Zackrisson et al. d}
Zackrisson, E, Bergvall, N., Leitet, E. 2008, {\it ApJL}, 676, 9 
\bibitem{Ferland et al.}
Ferland, G. J., Korista, K. T., Verner, D.A., Ferguson, J.W., Kingdon, J.B., Verner, E.M. 1998, {\it PASP}, 110, 761
\bibitem{Zackrisson et al. a}
Zackrisson, E., Bergvall, N., Olofsson, K., \& Siebert, A. 2001, {\it A\&A}, 375, 814
\end{thebibliography}
\end{document}